\documentclass[epj,referee]{svjour}
\usepackage{graphics}
\usepackage{amssymb}

\begin{document}

\title{Low energy phonon spectrum and its parameterization\
in pure KTaO$_3$ below 80 K}

\author{E. Farhi\inst{1}
\thanks{\emph{Present address:} Institut Laue Langevin, BP 156, 38042 Grenoble Cedex 9, France. \email{farhi@ill.fr}}
\and A.K. Tagantsev\inst{2}
\and R. Currat\inst{3}
\and B. Hehlen\inst{1}
\and E. Courtens\inst{1}
\and L.A. Boatner\inst{4}
} 

\institute{
Laboratoire des Verres, UMR 5587 CNRS, Universit\'{e} Montpellier 2, F-34095 Montpellier, France
\and Laboratoire de C\'eramique, EPFL, CH-1015 Lausanne, Switzerland
\and  Institut Laue Langevin, BP 156, 38042 Grenoble Cedex 9, France
\and  Solid State Division, Oak Ridge National Laboratory, Oak Ridge, TN 37831-6056, USA
} 

\date{Received: September 20, 1999 / Revised version: \today}


\abstract{
High resolution data on low energy phonon branches (acoustic  and soft optic)
along the three principal symmetry axes in pure KTaO$_3$ were obtained by cold
neutron inelastic scattering between 10 and 80 K. Additional  off-principal
axis measurements were performed to characterize the dispersion anisotropy
(away from the $<$100$>$ and $<$110$>$ axes). The parameters of the  phenomenological model
proposed by Vaks \cite{Vaks73} are refined in order to successfully
describe the experimental low temperature ($10 < T < 100$ K) dispersion
curves,
over an appreciable reciprocal space volume around the zone center ($|\vec{q}| < 0.25$ rlu). The refined
model, which involves only 4 temperature-independent adjustable parameters, is
intended to serve as a basis for quantitative computations of multiphonon processes.
\PACS{
	{77.84.Dy}{Niobates, titanates, tantalates, PZT ceramics, etc.} \and 
	{63.20.-e}{Phonons in crystal lattices} \and
	{62.20.Dj}{Phonon states and bands, normal modes, and phonon dispersion } \and
	{78.70.Nx}{Neutron inelastic scattering } 
}
} 

\maketitle



\section{Introduction}

The last decade has seen a renewal of interest in the paraelectric perovskite
crystals \cite{Kurtz63,Muller79} potassium tantalate, KTaO$_3$, and strontium
titanate, SrTiO$_3$. Both exhibit a large dielectric constant increase on
cooling connected with the softening of a transverse optic zone-center mode
(TO-mode). The softening of the TO-mode frequency $\Omega_{sm}$ is however
incomplete owing to zero-point fluctuations and both crystals remain {\it
paraelectric} down to the lowest temperatures ($T$). These materials,
alternatively called incipient ferroelectrics or quantum paraelectrics, exhibit
a wide range of anomalous thermal properties associated with the low frequency
TO-mode and its mixing at finite wavevector with the transverse acoustic (TA)
degrees of freedom \cite{Axe70,Hehlen98}.  Of particular interest is the
observation of an extra Brillouin scattering doublet with a sound-like
dispersion relation, originally attributed to {\it second sound}
\cite{Hehlen95.2,Hehlen95,Courtens96}. This new excitation appears on top of
the broad quasi-elastic central peak first reported by Lyons and Fleury
\cite{Lyons76}. Understanding these low-$T$ features requires a complete and
accurate knowledge of the low-frequency phonon dispersions not only along the
principal axes but also in non-symmetry directions.

The case of SrTiO$_3$ is more complicated than that of KTaO$_3$, because
SrTiO$_3$ undergoes the well-know cubic-to-tetragonal
antiferrodistortive structural transition at $T_a \sim 105$ K
\cite{Unoki67,Thomas68,Yamada69,Feder70}, below which the unit cell is doubled.
Owing to this, the crystal in the distorted phase ($T < T_a$) can form two
types of domains: three {\it tetragonal} domains with the $c$-axis parallel to
either one of the three $<$100$>$ cubic directions and, in each case, two {\it
antiphase} sub-domains corresponding to opposite rotational amplitudes for the
oxygen octahedra. While the tetragonal domains can be aligned by application of
uniaxial pressure \cite{Hehlen98,Hehlen95.2}, the antiphase domain structure is
more difficult to control and there is growing evidence that it may give rise
to anomalous spectral features \cite{Arzel99}.

In addition to these experimental difficulties, the modelisation of the phonon
spectra in tetragonal SrTiO$_3$ is intrinsically more complex due to the
presence of the three structural soft modes originating from the cubic-phase
R$_{25}$ zone-corner mode. For all these reasons, it appeared preferable to
first investigate in detail the ferroelectric TO-mode and acoustic mode
dispersions in KTaO$_3$ where the quantum paraelectric state is not perturbed
by the presence of other low-lying optic branches.

Information on the low frequency excitations in KTaO$_3$, is available from
scattering experiments such as, Raman \cite{Fleury67}, hyper-Raman
\cite{Inoue85,Vogt90}, Brillouin  \cite{Hehlen95,Courtens96,Lyons76}, and
neutron \linebreak scattering \cite{Axe70,Shirane67,Harada70,Comes72,Perry89}, and from
infrared spectroscopy \cite{Grenier89,Jandl91}. Indirect information is also
available from dielectric \cite{Rytz80} and thermal \cite{Salce94}
measurements. Neutron-scattering studies have revealed a strong anisotropy in
the TA-mode dispersion, suggesting an anisotropic TA-TO interaction mechanism
\cite{Axe70,Comes72}. Perry {\it et al.} \cite{Perry89} measured the complete set of phonon
dispersions along the three principal directions at several temperatures
between 4 $K$ and 1200 $K$. The data were analyzed using a self-consistent
anharmonic  shell-model taking account of the anisotropic non-linear
polarizability of the oxygen O$^{2-}$ ion \cite{Perry89,Migoni76,Bilz87}. 

In the present paper we are actually interested in a simpler phenomenological
approach limited to the low energy branches near the center of the Brillouin
zone (BZ). For this purpose an expansion of the phonon eigenfrequencies
($\omega$) in powers of the wavevector ($\vec{q}$) can be used, and such a
theory has already been developed by Vaks \cite{Vaks68,Vaks73}. We shall
show that the model parameters estimated by Vaks \cite{Vaks73} do not describe
the low-energy low-$T$ phonon dispersion curves in KTaO$_3$ with the degree
of accuracy needed for computational purposes. In order to reliably
evaluate the expansion parameters in the model, more extensive and more
accurate data are in fact necessary. Although high resolution data from cold
neutron measurements are available for Li and Nb doped KTaO$_3$
\cite{Hennion94,Hennion96,Fontana93,Foussadier96},  no such data exist for the pure crystal. For
this reason additional neutron scattering measurements were undertaken as
described in Section \ref{Expe} below. The results of these measurements are
presented in Section \ref{Measures}, while Section \ref{Vaks} describes their
parameterization with the Vaks model. The quantitative low-$T$ multiphonon
computations  based on the Vaks-model parameterization will be published
separately \cite{Farhi00,Farhi98}.

\section{Experimental conditions and data analysis}
\label{Expe}

\subsection{Sample}

The Ultra High Purity (UHP) crystal used for this experiment was  grown by one
of us at the Oak Ridge National Laboratory using the "modified spontaneous
nucleation" method in a slowly cooled flux \cite{Thomson82}. This technique
yields  single crystals with blue semiconducting parts surrounded by colorless
UHP-material. The sample used here is a 2.5 $\times$ 1 $\times$ 0.2 cm$^3$
polished colorless plate, with faces normal to the $<$100$>$ principal
axes. The effective mosaic of the plate, originally below 5', increased to
about 20' in the course of the experiments, possibly owing to repeated  thermal
cycling. The crystal was mounted in a $^4$He cryostat, with its vertical axis
along a [001] or [0$\overline{1}$1] direction.  It should be remarked that the
thickness of 0.2 cm is near optimal for cold neutron work due to the large
absorption cross-section of Ta ($\sigma_a$(natural Ta) $\simeq 40$ $bn$ at 6
meV).

\subsection{Scattering technique}

The experiments were performed on the cold neutron three axis spectrometer
IN14 at the Institut Laue Langevin, Grenoble, France. The instrument is equipped with a pyrolitic
graphite (002) flat monochromator, and the scattered neutron wavevector
$k_f$ was held fixed at either 1.785 \AA$^{-1}$ or 2 \AA$^{-1}$. Collimation was achieved
with 40' and 60' Soller's, before and after the sample, respectively. As
analyzer, a horizontally focusing silicon (111) crystal was used in order
to suppress the  second order neutrons ($2k_f$). The horizontal analyzer curvature
broadens the elastic \linebreak $\vec{q}$-resolution,  but does produce a significant
intensity gain, with only slight effects on phonon line-widths. The
instrumental ($\vec{q},\omega$) elastic resolution ellipsoid was mapped out
using the KTaO$_3$ (200) Bragg reflection. It  is well represented by a
4-dimensional gaussian distribution with widths (hwhm) $\Delta q_x =$ 0.017
rlu, $\Delta q_y =$ 0.013 rlu, $\Delta q_z =$ 0.05 rlu, and $\Delta E =$
0.17 meV, where the indices $x$, $y$ and $z$ refer to longitudinal [100],
transverse in-plane [010] and vertical [001] axes respectively. The
reciprocal lattice \linebreak units (rlu's) are defined as $\xi = qa/2\pi$, where
$a=3.983$ \AA $\;$ is the lattice parameter of KTaO$_3$.

Excitations were measured on the Stokes side (neutron energy loss) over a
typical energy range from 0 up to a maximum of 8 meV. The phonon branches
along the 3 principal directions were scanned up to $\xi=$ 0.5 rlu with 
special attention given to the vicinity of the BZ center, $|\xi| < 0.2$
rlu. A total of approximately 150 high resolution inelastic scans was
obtained at 10, 24, 39, 60 and 80 K, amounting to an equivalent of 4 weeks
measurement time.

\subsection{4D-Monte Carlo adjustment}

\begin{figure}[ht]
\begin{center}
\includegraphics{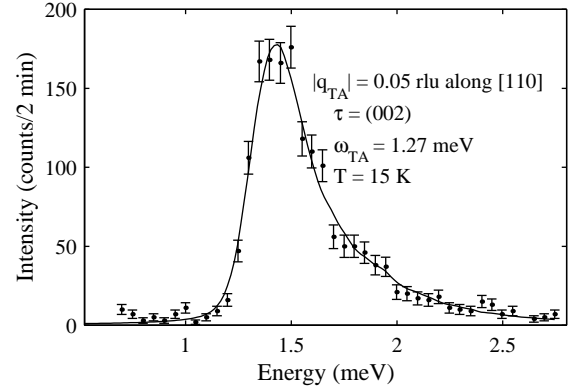}
\end{center}
\caption{\label{sv6624} $\omega$-scan : TA phonon in the [110] direction
($|\vec{q}|=$ 0.05 rlu), polarized along [001]; KTaO$_3$, $T =$ 15 K. The fit
is performed using a '4D' Monte Carlo convolution technique.}
\end{figure}

\begin{figure}[ht]
\begin{center}
\includegraphics{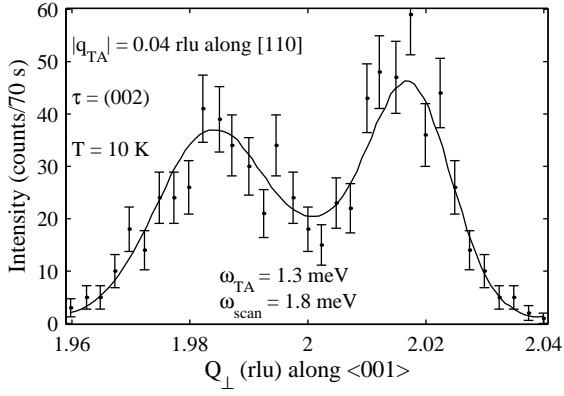}
\end{center}
\caption{\label{sv6647} $q_{\perp}$-scan across a TA 'valley' around $|\vec{q}|
=$ 0.04 rlu in the [110] direction; KTaO$_3$, $T = 10$ K. The fit is performed
using a '4D' Monte Carlo convolution technique.}
\end{figure}

The inelastic neutron scattering data treatment was performed using the
usual procedure in the case of anisotropic dispersion surfaces 
\cite{Hennion94,Hennion96}. First, the gaussian 4-\linebreak dimensional
$(\vec{q},\omega)$ modelization of the resolution ellipsoid was performed
using the Popovici method \cite{Popovici75} implemented in {\it Rescal}
\cite{Rescal}. The model ellipsoid was then convoluted with a locally
anisotropic quadratic expansion of the dispersion surface using a Monte
Carlo algorithm around a reference point $({\vec{\xi_0}},\omega_0)$. The
expansion is taken as
\begin{equation}
\label{valquad}
\omega^2(\vec{\xi} - \vec{\xi_0}) = (\omega_0 + c_{//} \xi_{//})^2 +
(c_{\perp,1} \xi_{\perp,1})^2 + (c_{\perp,2} \xi_{\perp,2})^2 \; ,
\end{equation}
where $\xi_{//}$ is the projection of the reduced wavevector shift,
$\vec{\xi} - \vec{\xi_0}$, along a direction set by the symmetry of the
dispersion surface at the reference point $\vec{\xi_0}$, while
$\xi_{\perp,1}$ and $\xi_{\perp,2}$ are the components of the shift  normal
to this direction. The $c$'s are appropriate coefficients defining the
local curvatures of the dispersion surface, whose numerical values are
obtained iteratively. Expression (1) is of course modified appropriately if
the reference point is at $\vec{\xi_0} = 0$, in which case $\omega_0$ is a
local minimum for the TO-mode. Phonon line-shapes are taken as damped
harmonic oscillators \cite{Dorner97} with eigenfrequency $\omega_j$ and damping
coefficient $\Gamma_j$,
\begin{equation}
S_j(\vec{q},\omega) = \frac{1}{\pi} \frac{[1+n(\omega)]\omega
\Gamma_j(\vec{q})}{(\omega^2-\omega_j^2(\vec{q}))^2 +
(\Gamma_j(\vec{q})\omega)^2} \; .
\label{dho}
\end{equation}
The actual fit is obtained  using a least square convergence criterion
implemented in a Marquardt-Levenberg gradient algorithm \cite{MFit}.  A
typical energy scan along a principal axis is illustrated in Fig.
\ref{sv6624}. A typical off-axis wavevector scan, perpendicular to a
principal direction, is shown in Fig. \ref{sv6647}. As already remarked in
\cite{Comes72,Perry89}, the asymmetrical line profile seen in Fig.
\ref{sv6624}, is fully accounted for by the finite wavevector
resolution combined with the steep and anisotropic dispersion of the
TA-mode.

\section{Experimental results}
\label{Measures}

\subsection{Zone-center soft TO-mode}

\begin{figure}[ht]
\begin{center}
\includegraphics{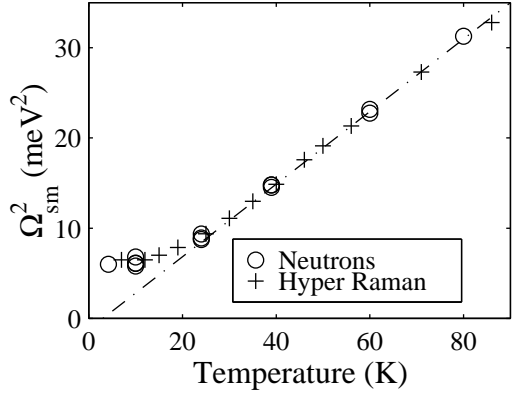}
\end{center}
\caption{\label{wTOq0} KTaO$_3$: soft TO-mode squared frequency at the zone
center compared to previous hyper-Raman measurements \cite{Vogt96}. The mean
energy uncertainty for the neutron data is $\Delta E \sim$ 0.1 meV.}
\end{figure}

\begin{figure}[ht]
\begin{center}
\includegraphics{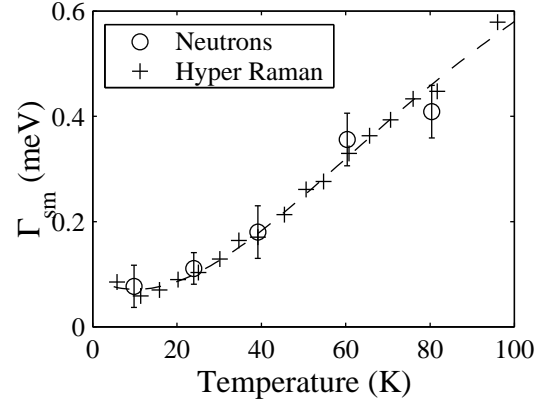}
\end{center}
\caption{\label{gTOq0} KTaO$_3$: Damping coefficient (half width) of soft
zone center TO-mode compared to previous hyper-Raman measurements
\cite{Vogt96}. The mean uncertainty for the neutron data is $\Delta \Gamma
\sim $ 0.1 meV.}
\end{figure}

The frequency $\Omega_{sm}$ and damping $\Gamma_{sm}$ of the soft  $F_{1u}$
($\Gamma_{15}$) TO-mode at the BZ-center were obtained as  a function of
temperature. As shown in Fig. \ref{wTOq0}, the squared TO-frequency follows a
standard Curie-Weiss law above $ \sim$ 30 K, and saturates at a finite
value $\Omega_{sm}^2 \sim$ 6 meV$^2$ below 10 K. This is in perfect
agreement with previous hyper-Raman determinations
\cite{Inoue85,Vogt90,Vogt96}. For the mode-damping coefficient (Fig. \ref{gTOq0}),
the present neutron-scattering measurements also agree perfectly with
recent hyper-Raman data on a sample of similar quality \cite{Vogt96}.

\subsection{Low energy phonon-dispersion curves}

\begin{figure*}[h]
\begin{center}
\includegraphics{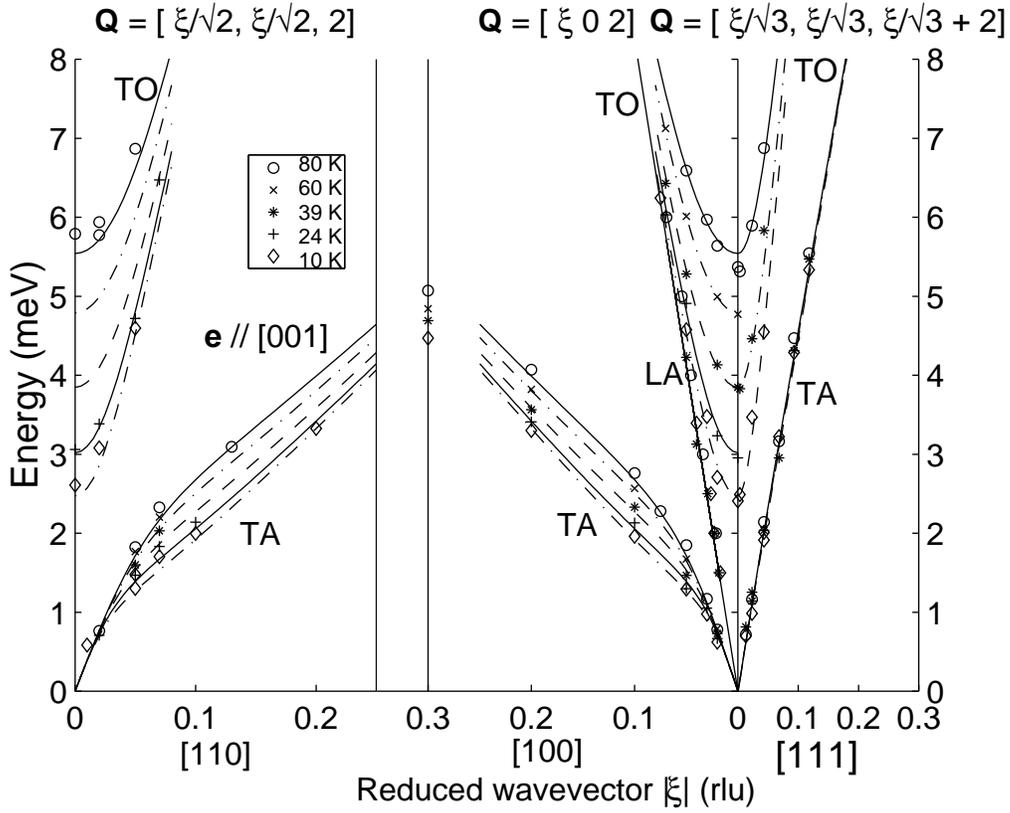}
\end{center}
\caption{\label{DC} Dispersion curves of KTaO$_3$ low energy phonons between
10 and 80 K along the high symmetry axes.
The mean energy uncertainty is 0.1 meV. The LA mode is measured at $\vec{Q} =
[$0, 0, 2+$\xi]$. The lines correspond to calculations using the
Vaks parameterization presented in this work.}
\end{figure*}

Fig. \ref{DC} shows the phonon-dispersion branches measured in the high
symmetry directions. These data complement the previous results from Refs.
\cite{Axe70,Comes72,Perry89}. The transverse \linebreak acoustic phonons measured at
$\vec{Q}=[\xi,0,2]$ and \linebreak $[\xi/\sqrt{2}, \xi/\sqrt{2},2]$, are
polarized along the [001] axis. They both correspond to the elastic
constant $C_{44}$ at small $\xi$ and hence have the same dispersion over
the central region of the BZ. The coupling between the TA and TO-modes
lowers the acoustic phonon frequency as the TO-mode softens on decreasing
$T$ \cite{Axe70,Perry89}. This feature affects all TA branches polarized
along $<$100$>$ directions. Hence it is not seen on the TA-branch
propagating in the [111] direction, nor on the TA-branch propagating along
[110] and polarized along [1$\overline{1}$0] (not shown in Fig. \ref{DC}).

\begin{figure}[ht]
\begin{center}
\includegraphics{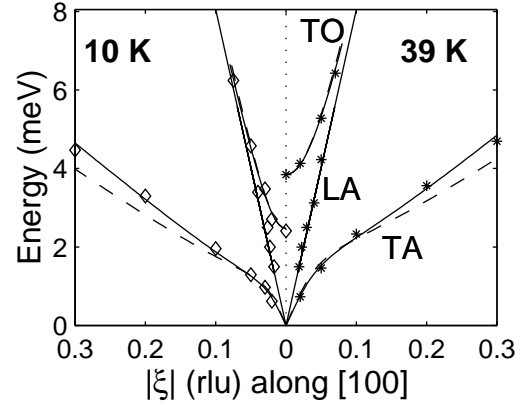}
\end{center}
\caption{\label{vaksold} KTaO$_3$: Dispersion curves measured at 10 and 39 K
 in the [100] direction;  the original 
Vaks parameterization \cite{Vaks73} is shown as dashed lines. The modified
Vaks model (this work) is shown as solid lines.}
\end{figure}

The slopes of the TA-dispersion curves in the vicinity of the BZ-center are
in good agreement with the elastic constants that are known from neutron
and Brillouin scattering  \cite{Hehlen95,Perry89,Farhi98}. In contrast to
the case of the TO-mode, no measurable damping could be extracted from the
TA and LA-mode experimental line-shapes. This sets an upper limit for
$\Gamma_{TA}$ and $\Gamma_{LA}$ at about 0.05 meV. Finally, as seen in
Fig.  \ref{vaksold}, the original Vaks model parameters \cite{Vaks73} do
not describe satisfactorily low-$T$ TA-modes along high symmetry axes for
$|\vec{q}| \gtrsim$ 0.1 rlu ($\omega \gtrsim$ 2 meV).

\subsection{Anisotropy of the spectrum}
\label{ExpeRes:Anis}

In order to characterize the phonon-spectrum anisotropy, some off-principal axis
scans were performed. Starting from a reciprocal space position on a high
symmetry axis, \linebreak $\vec{q}$-scans perpendicular to this axis have been
performed. A typical $\vec{q}$-scan at constant $\omega$ is shown in Fig.
\ref{sv6647}. Some $\omega$-scans at constant off-axis $\vec{q}$ were also
performed. Com\`es and Shirane first studied this anisotropy at 20 K and
300 K by mapping the LA, TA and TO-mode dispersions in the vicinity of the
[100] axis. They found that the TA and TO-mode dispersions form valleys
that become deeper as $T$ decreases, while the LA-mode exhibits a weak
anisotropy. They ascribed the characteristic diffuse streaks observed by
X-ray scattering \cite{Comes68} to the strong anisotropy of the TA-mode dispersion.

\begin{figure}[ht]
\begin{center}
\includegraphics{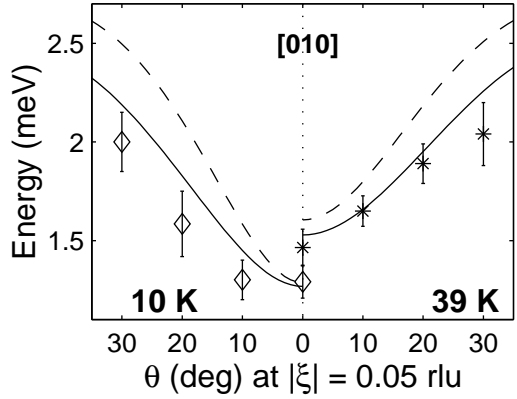}
\end{center}
\caption{\label{vakstheta} Measured dispersion curves in KTaO$_3$ at 10 and
39 K near the [010] direction in the (001) scattering plane: $\vec{Q}$ =
[(2+$\xi\sin\theta$), ($\xi\cos\theta$), 0]; The dashed and solid lines
refer to the original \cite{Vaks73} and modified (this work) Vaks parameters.}
\end{figure}

\begin{figure}[ht]
\begin{center}
\includegraphics{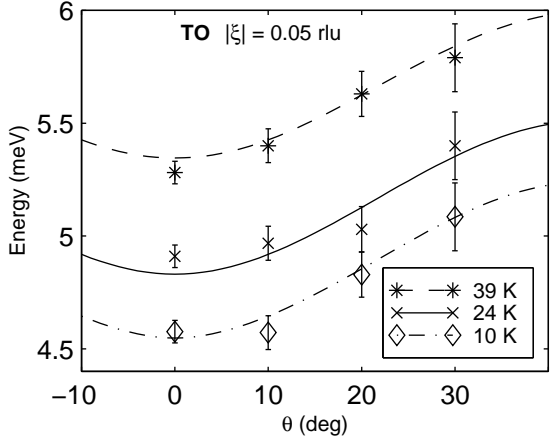}
\end{center}
\caption{\label{aTOq5} Anisotropy of the TO phonon branch near the [010]
symmetry axis in KTaO$_3$: $\vec{Q}$ = [(2+$\xi\sin\theta$),
($\xi\cos\theta$), 0]. The lines are calculated with the modified Vaks
parameters.}
\end{figure}

Our off-axis scans confirm this behaviour for the TA and TO-branches around
the [100]  and [110] directions (Figs. \ref{vakstheta} and \ref{aTOq5}).
The energy accuracy of $\sim$ 0.1 meV  which is achieved on-axis
deteriorates slightly  as one moves away from the high symmetry direction.
It is clear from Fig. \ref{vakstheta} that the Vaks model using the
original KTaO$_3$ parameter set \cite{Vaks73} is not satisfactory for an
accurate parameterization of the dispersion surfaces away from the
principal axes, even at $\vec{q}$ values as low as 0.05 rlu.

\section{Parameterization of the phonon spectrum}
\label{Vaks}

Two alternate approaches have been used to describe the phonon spectrum of
KTaO$_3$. First, the phenomenological description due to Vaks
\cite{Vaks68,Vaks73} consists in expanding the lattice Hamiltonian in powers of
$\vec{q}$, taking only into \linebreak account the low lying phonon branches, and the
interactions among them, as allowed by symmetry. The expansion, limited to the
5 lowest-energy excitations (two TA, one LA, and two TO-modes), is actually
equivalent to a Ginzburg-Landau free energy expansion with respect to
dielectric polarization and elastic deformations. Similar ideas have been
discussed by Axe {\it et al.} \cite{Axe70} in order to account for the strong
$T$-dependent depression in the TA-mode dispersion owing to the repulsion with
the softening TO-mode. In the Ginzburg-Landau formalism, this effect is described as a symmetry-allowed
interaction between the shear strain and the {\it gradient} of the electric
polarization.

A more ambitious approach was developed by Migoni {\it et al.} \cite{Migoni76}.
This is a microscopic model which aims at \linebreak describing the complete phonon
spectrum across the entire BZ, and  which accounts for the softening of the
TO-mode through the non-linear polarizability of the O$^{2-}$ ion. Perry {\it
et al.} \cite{Perry89} used this model to describe the phonon branches of
KTaO$_3$ along the high symmetry directions. Their extensive neutron
measurements allowed to determine the full set of 15 model parameters. However,
the resolution of the experiments, and thus the accuracy of the resulting
model, are not better than $\sim$0.5 meV.

At low temperatures, the thermodynamic and phonon-kinetic properties of
KTaO$_3$ are controlled by the 5 lowest modes included in the Vaks model
\cite{Vaks68,Vaks73}. Since our aim is not to explain the origin of the soft
mode, but rather to use the phonon spectrum to account for the low temperature
anomalies observed in the quantum paraelectric state, the Vaks
parameterization is a very convenient one. It includes 8 adjustable parameters,
but 4 of these are well determined from light-scattering data. As the physical
\linebreak basis of the model is only detailed in Ref. \cite{Vaks73}, we summarize it
below, specializing to the cubic case which is of interest here. We shall then
refine the parameter determination in order to obtain a reliable description of
low-lying dispersion surfaces.

\subsection{Effective Hamiltonian}
\label{sec41effham}
One considers first the $harmonic$ part of the lattice Hamiltonian $\cal H$ of
a cubic perovskite crystal. Let $\vec{u}_i(\vec{q})$ be the spatial Fourier
transform of the displacement from equilibrium for the $i$th atom in the unit
cell. The mass of this atom is denoted $m_i$ and its effective charge, $e_i$.
One finds \cite{Vaks73,Born54,Kwok66}
\begin{eqnarray}
\label{C3HamTot}
\index{Hamiltonien}
{\cal H} & = & \frac{1}{2} \sum_{i,\vec{q}} m_i |\vec{ \dot{u_i}} (\vec{q})| ^2 \nonumber \\
& + & \frac{1}{2} \sum_{i_1,i_2, \vec{q} } \vec{ u}_{i_1} (\vec{q}) \hat{ \Phi} _{i_1,i_2} (\vec{q}) \vec{u}_{i_2} (-\vec{q})  \nonumber \\ 
& - & \sum_{i,\vec{q}} e_i \vec{ u_i} (\vec{q}) \vec{E} (-\vec{q}) + \frac{ \epsilon_\infty }{2} \sum_{\vec{q}} |E(\vec{q})| ^2,
\end{eqnarray}
where $\epsilon_\infty $ is the high frequency dielectric permittivity of the
crystal. The tensor $\hat{\Phi}$ accounts for the "short-range" part of the
potential energy, {\it i.e.} for the contribution of the short-range forces
plus the analytical (in $\vec{q}$, at $\vec{q} \rightarrow 0$) part of the
dipole-dipole interaction. The non-analytical part is taken into account by
introducing the Fourier transform of the macroscopic electric field $\vec{E}
(\vec{q})$, which is given by Poisson's equation,
\begin{equation}
\vec{q} \cdot \left[ \epsilon_\infty \vec{E} (\vec{q}) +
\vec{P} (\vec{q}) \right] = 0 \; .
\label{Poisson}
\end{equation}
The Fourier transform of the macroscopic electric polarization
is defined by
\begin{equation}
\vec{P} (\vec{q}) = \frac{1}{v_c} \sum_{i} e_i \vec{ u_i} (\vec{q}) \; ,
\label{Polarization}
\end{equation}
where $v_c=a^3$ is the unit-cell volume.

Using a routine procedure, Eqs. (\ref{C3HamTot}), (\ref{Poisson}), and
(\ref{Polarization}) give an effective Hamiltonian that can be diagonalized in
terms of normal phonon coordinates. However, following Vaks
\cite{Vaks68,Vaks73} the most convenient form of the Hamiltonian  for the
present purpose is obtained using a linear transformation of the displacements
$\vec{u}_i(\vec{q})$ which diagonalizes only the short-range part of $\cal H$
in the limit $\vec{q} \rightarrow 0$. One selects a particular $\vec{u}$ to
describe the acoustic modes by
\begin{equation}
\vec{u}_1(\vec{q}) = \frac{1}{\sqrt{m_1}} \vec{u}(\vec{q}) \; .
\label{Acoustic6}
\end{equation}
One defines then the optical displacements $\vec{x}_b(\vec{q})$ by
\begin{equation}
u_{i,\alpha}(\vec{q}) - u_{1,\alpha}(\vec{q})
 = \frac{1}{\sqrt{m_i}} \sum_{\beta,b=1}^{4} L_{i\alpha,b\beta}
x_{b,\beta}(\vec{q}) \; ,
\label{Optic7}
\end{equation}
where $\alpha$ and $\beta$ are cartesian indices. In writing (\ref{Optic7}),
one takes into account that the short-range part of the lattice Hamiltonian of
a cubic perovskite at small $q$ contains 4 sets of 3-fold-degenerate optic
modes that correspond to the same vector irreducible representation. The
subscript $b$ indexes these sets. With the new variables, Eqs.
(\ref{C3HamTot}), (\ref{Poisson}), and (\ref{Polarization}) give:
\begin{eqnarray}
{\cal H} & = & \frac{1}{2} \sum_{\vec{q}} [\vec{ \dot{u}}(-\vec{q}) \vec{\dot{u}}(\vec{q}) + \vec{u} (-\vec{q}) \hat{A}(\vec{q}) \vec{u} (\vec{q})  \nonumber \\
& + & \sum_b \{ \vec{\dot{x}}_b(-\vec{q}) \vec{\dot{x}}_b(\vec{q}) +
\lambda_b \vec{x}_b(-\vec{q}) \vec{x}_b(\vec{q})\nonumber \\
& + & 2\vec{u} (-\vec{q}) \hat{V}_b(\vec{q}) \vec{x}_b(\vec{q}) \}
 \nonumber \\
& + &  \sum_{b,b'} \vec{x}_b(-\vec{q}) \{\hat{S}_{b,b'} (\vec{q}) + \hat{\Phi}_{b,b'}^E (\vec{q}) \} \vec{x}_{b'}(\vec{q}) ] \; ,
\label{Hharm}
\end{eqnarray}
where $\lambda_b$ are the restoring short-range force constants of the optic
modes. The tensors $\hat{A}$, $\hat{V}_b$, and $\hat{S}_{b,b'}$ correspond to
the contributions of the  short-range interactions. These analytically go to
zero as $q^2$ for $\vec{q} \rightarrow 0$. The tensor $\hat{\Phi}_{b,b'}^E $ is
non-analytic at $\vec{q} \rightarrow 0$. It represents the contribution from
the long-range part of the dipole-dipole interaction,
\begin{equation}
\label{Coulomb}
(\Phi_{b,b'}^E)_{\alpha\beta} = \sum_{\alpha', \beta'}\left(
\frac{ z_{b, \alpha \alpha '} z_{b', \beta \beta' } }{\epsilon_\infty a^3}
\right)
n_{\alpha'} n_{\beta'} ,
\end{equation}
where $\vec{n} = \vec{q}/|\vec{q}|$, and $z_{b,\alpha \alpha'} = \sum_{i} e_i
L_{i \alpha,b \alpha'} / \sqrt{m_i}$ is the effective charge tensor for the
$b$-set of optic modes in the limit $\vec{q} \rightarrow 0$, which in view of
the cubic symmetry is $\propto \delta_{\alpha \alpha'}$  \cite{Vaks73}.

One is only interested in the low frequency part of the spectrum. Following
Vaks \cite{Vaks68,Vaks73}, one first diagonalizes the optical part of
(\ref{Hharm}) in the limit $\vec{q} \rightarrow 0$, keeping only the terms
containing the restoring short-range force constants of the optic modes ,
$\lambda_b$, and the non-analytical term $\hat{\Phi}_{b,b'}^E $. One finds that
the latter lifts the degeneracy of the optical modes, splitting each set into a
pair of degenerate TO-modes, whose frequency squared is equal to $\lambda_b$,
and a longitudinal (LO) mode.  One pair of TO-modes in the set $b$ corresponds
to the soft mode, which has a frequency $\Omega_{sm}$ much lower than all other
optic modes. It is associated to the hardest LO mode \cite{Zhong94}, labelled
LO$_4$ in the Perry {\it et al.} denomination \cite{Perry89}. We denote the
corresponding $\lambda_b$ by $\lambda$ and the corresponding optical
displacement vector $\vec{x}_b$ by $\vec{x}$. The anharmonic interactions, not
included in \ref{C3HamTot}, are taken into account with a standard {\it
quasi-harmonic} approximation. The frequency of the soft mode becomes then
temperature dependent through the $T$-dependence of $\lambda = \Omega^2_{sm}$.
In futher considerations, only the subset consisting of the two soft TO-modes
and of the 3 acoustic modes is kept. One neglects the other optical modes as
well as their interactions with this subset. As shown by Vaks
\cite{Vaks68,Vaks73}, the accuracy of this approximation for the description of
a phonon of frequency $\omega$ is controlled by the small parameter
${\omega^2}/{\Omega^2_{op}}$, where $\Omega_{op}$ is a typical hard optical
mode frequency such as the hard LO$_4$ mode. Hence for the
part of the spectrum that we are interested in ($\omega\simeq\Omega_{sm}$), it
results from the Lyddane-Sachs-Teller relation that this parameter is of order
$\epsilon_\infty/\epsilon_{o}$, where $\epsilon_{o}$ is the static dielectric
permittivity of the crystal.

Technically, the separation of the Hamiltonian containing only the 5 modes of
interest is done to within the aforementioned accuracy by considering the
problem in a rotated reference frame $\vec{X^\prime} \equiv (X^\prime_1,
X^\prime_2, X^\prime_3)$, rather than in the usual cubic frame  $\vec{X} \equiv
(X_1, X_2, X_3)$.  The axis $X^\prime_3 $ is taken parallel to $\vec{q}$.
Specifically, the rotated frame can be defined by the relation $\vec{X}^\prime
= \hat{M} \vec{X} $, where the matrix $\hat{M}$ is given as a function of the
direction of $\vec{q}$ ($\vec{n} = \vec{q}/q$) by
\begin{eqnarray}
\label{C3PolarMat}
M(\vec{n}) = 
\left(
\begin{array}{ccc}
0 		& n_\perp 		& n_1 \\
-n_3/n_\perp	& -n_1 n_2/n_\perp 	& n_2 \\
n_2/n_\perp	& -n_1 n_3/n_\perp 	& n_3 
\end{array}
\right),
\end{eqnarray}
where $\vec{n}$ is written in the cubic reference frame $\vec{X}$, and
$n_\perp^2 = 1 - n_1^2$.

One sees that in the rotated frame the components of the acoustic and optical
bare mode displacements (as defined in (\ref{Acoustic6}) and (\ref{Optic7})), associated with
the 5 modes of interest are just $x_1, x_2, u_1, u_2$, and $u_3$, while $x_3$
is essentially the displacement of the LO$_4$ mode which has been eliminated.
Finally, the 5-mode Hamiltonian can be written
\cite{Vaks68,Vaks73}:
\begin{eqnarray}
\label{C3HamDiag}
{\cal H}^{(5)} & = & \frac{1}{2} \sum_{\vec{q}} [ \dot{\vec{u}}_{-q} \dot{\vec{u}}_{q} + \vec{u}_{-q} {\hat A}(\vec{q}) \vec{u}_{q} + \dot{\vec{x}}_{-q} \dot{\vec{x}}_{q} \nonumber  \\
	& + & \lambda \vec{x}_{-q} \vec{x}_{q} + \vec{x}_{-q} {\hat S}(\vec{q}) \vec{x}_{q} \\
	& + & 2 \vec{u}_{-q} \hat{V}(\vec{q})\vec{x}_{q}] \nonumber\; .
\end{eqnarray}
The components of the  tensors $\hat{A}$, $\hat{S}$ and $\hat{V}$ that give
non-vanishing contributions in the subspace ($x_1, x_2, u_1,
u_2, u_3$) in the small-$q$ limit are
\begin{eqnarray}
\label{C3ASV}
\hat{A} & = & q^2(A_l g^l + A_t g^t + A_a g^a),  \nonumber \\
\hat{S} & = & q^2(S_t g^t + S_a g^a),  \\
\hat{V} & = & q^2(V_t g^t + V_a g^a),  \nonumber
\end{eqnarray}
with
\begin{eqnarray}
g^l_{\alpha \beta} & = & n_\alpha n_\beta, \nonumber \\
g^t_{\alpha \beta} & = & \delta_{\alpha \beta} - n_\alpha n_\beta,  \label{C3g} \\
g^a_{\alpha \beta} & = & \gamma_{\alpha \beta \gamma \delta} n_\gamma
n_\delta \nonumber.
\end{eqnarray}
The tensor $\gamma_{\alpha \beta \gamma \delta}$ is defined, in the cubic
frame, by $\gamma_{\alpha \beta \gamma \delta} = 1$ if $\alpha =\beta =\gamma
=\delta$, and $\gamma_{\alpha \beta \gamma \delta}=0$ otherwise. In that frame,
$\hat{A}$, $\hat{S}$, and $\hat{V}$ stand for acoustic, optic, and
acoustic-optic coupling tensors respectively, split into transverse $g^t$,
longitudinal $g^l$, and anisotropic $g^a$ parts, as seen in (\ref{C3ASV}). The
coefficients $S_l$ and $V_l$ do not appear in (\ref{C3ASV}) since $\vec{x}$ is
taken as purely transverse.

Since the kinetic energy terms are already diagonal, the phonon-dispersion
curves for the 5 branches are given by the square roots of the
eigenvalues of the following matrix:

\begin{eqnarray}
{\cal H}^{(5)}(\vec{q}) & = & {\cal H}_{is}(|\vec{q}|,\lambda,S_t,A_t,V_t,A_l) \nonumber \\
	& + & {\cal H}_{anis}(\vec{q},S_a,A_a,V_a) \; .
\label{HamFin}
\end{eqnarray}
The isotropic matrix, ${\cal H}_{is}$, is
\begin{eqnarray}
\left(
\begin{array}{ccccc}
\lambda + S_t q^2 & 0 & V_t q^2 & 0 & 0 \\
0 & \lambda + S_t q^2 & 0 & V_t q^2 & 0 \\
V_t q^2 & 0 & A_t q^2 & 0 & 0 \\
0 & V_t q^2 & 0 & A_t q^2 & 0 \\
0 & 0 & 0 & 0 & A_l q^2 
\end{array}
\right),
\end{eqnarray}
and the anisotropic one, ${\cal H}_{anis}$, is
\begin{eqnarray}
q^2
\left(
\begin{array}{ccccc}
S_a h_{11} & S_a h_{12} & V_a h_{11} & V_a h_{12} & V_a h_{13} \\
S_a h_{12} & S_a h_{22} & V_a h_{12} & V_a h_{22} & V_a h_{23} \\
V_a h_{11} & V_a h_{12} & A_a h_{11} & A_a h_{12} & A_a h_{13} \\
V_a h_{12} & V_a h_{22} & A_a h_{12} & A_a h_{22} & A_a h_{23} \\
V_a h_{13} & V_a h_{23} & A_a h_{13} & A_a h_{23} & A_a h_{33} 
\end{array}
\right),
\end{eqnarray}
in which
\begin{eqnarray}
\begin{array}{lll}
h_{11} & = & \frac{2 n_2^2 n_3^2}{n_\perp^2} \; , \nonumber \\
h_{12} & = & \frac{n_1 n_2 n_3}{n_\perp^2} (n_3^2 - n_2^2) \; ,\nonumber \\
h_{22} & = & 2 n_1^2 (n_\perp^2 - \frac{n_2^2 n_3^2}{n_\perp^2}) \; , \nonumber\\
\end{array}
\begin{array}{lll}
h_{13} & = & \frac{n_2 n_3}{n_\perp} (n_2^2 - n_3^2) \; ,\nonumber \\
h_{23} & = & \frac{n_1}{n_\perp} (n_1^2 n_\perp^2 - n_2^4-n_3^4) \; , \\
h_{33} & = & n_1^4 + n_2^4 + n_3^4 \; . \nonumber
\end{array}
\end{eqnarray}
The eigenvalues $\omega_i(\vec{q})$ of ${\cal H}^{(5)}(\vec{q})$ are the
squared phonon frequencies, and the diagonalization matrix $L_{i,j}(\vec{q})$ reflects the
mode mixing. Thus, the renormalized acoustic $\vec{u}$ and optical $\vec{x}$ displacements in the canonical cubic axis system are obtained through a linear combination of bare mode displacements:
\begin{equation}
\label{C3getXU}
\vec{x}_{j,\vec{q}} = \sum_{j'_{opt}=1,2} \vec{M}_{j'} L_{j,\vec{q}}^{(j')} \mbox{ and } \vec{u}_{j,\vec{q}} = \sum_{j'_{ac}=3,4,5} \vec{M}_{j'-2} L_{j,\vec{q}}^{(j')},
\end{equation}
where $\vec{M}_{j}$ indicates the column $j$ in matrix $M$.

In the matrix ${\cal H}_{is}$, the parameters $S_t$, $A_t$, and
$A_l$ appear as squared dispersion-curve slopes whereas $V_t$ is a TA-TO
coupling parameter, and $S_a$, $A_a$, and $V_a$ contribute to the off-principal
axis squared slopes.

\subsection{Adjustment of the phonon-dispersion surfaces}

We now use the above model to parameterize the measured phonon spectra. Since
the parameter set given by Vaks for KTaO$_3$ (see Table \ref{C3ParFit} and
Figures \ref{vaksold} and \ref{vakstheta}) leads to typical deviations of 0.5
meV or more with respect to our inelastic neutron scattering measurements, it is
necessary to recompute the parameter values. The model contains 8 parameters:
$S_t$, $A_t$, $V_t$, $S_a$, $A_a$, $V_a$, $A_l$,  and $\lambda =
\Omega_{sm}^2$, where only the latter is appreciably $T$-dependent. Three
parameters can be expressed in terms of the elastic constants $C_{ij}$ and the
density $\rho = 6.967$ g/cm$^3$ of the crystal \cite{Vaks68,Vaks73}:
\begin{eqnarray}
A_l & = & (C_{12}+2 C_{44})/{\rho} \; , \nonumber \\
A_t & = & C_{44}/{\rho} \; , \\
A_a & = & (C_{11}-C_{12}-2 C_{44})/{\rho} \; . \nonumber
\end{eqnarray}
The values of the elastic constants are known from Brillouin-scattering
experiments \cite{Hehlen95,Farhi98}: $C_{11} =4.344 \pm 0.049$, $C_{12} =1.027
\pm 0.038$, and $C_{44}= 0.990 \pm 0.020$ in 10$^{11}$ N/m$^2$. The
$T$-dependent values of $\lambda$ are available from the hyper-Raman
measurements \cite{Inoue85,Vogt90,Vogt96}. The values used presently are
given in Table \ref{C3lambda}. Thus, there remain only 4 adjustable parameters
at our disposal, namely $S_t$, $V_t$, $S_a$, and $V_a$. We adjusted  our data
on the 5 lowest frequency modes for the $<$100$>$, $<$110$>$, and $<$111$>$
directions to the square roots of the eigenvalues of the matrix (\ref{HamFin}).
The fit was performed with a Marquadt-Levenberg algorithm and a least square
criterion, for the temperatures, 10, 24, 39, 60, and 80 K, keeping the
parameters temperature independent. The overall correlation coefficient when
using 151 neutron and Brillouin measurements is $r^2 = 0.995$, and the
determined model parameters are given in Table \ref{C3ParFit}. This table also
contains the sets of parameters determined by previous workers
\cite{Axe70,Vaks68,Vaks73}. Appreciable differences in some of these can be
seen. The obtained parameterization of the spectrum has been tested by
comparing the phonon frequencies measured off principal directions with those
calculated from the parameters, as shown in Figs.\ref{vakstheta} and
\ref{aTOq5}. One sees that the model gives also a reasonable description for
the off-axis dispersion.

\begin{figure}[ht]
\begin{center}
\includegraphics{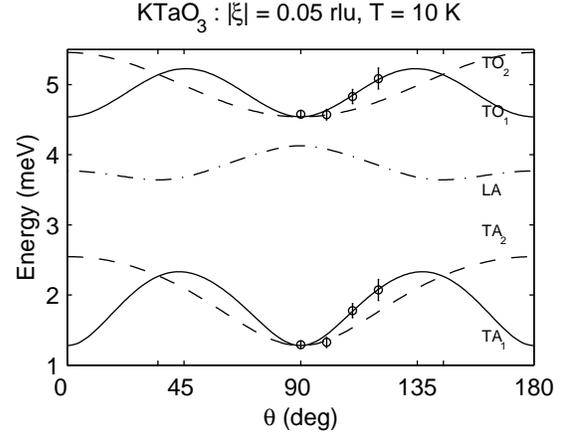}
\end{center}
\caption{\label{rot2d} KTaO$_3$: The low frequency phonon spectrum at $T =$ 10
K, for $\vec{q}$ = [($\xi\sin\theta$), ($\xi\cos\theta/\sqrt{2}$),
($\xi\cos\theta/\sqrt{2}$)] with $|\vec{\xi}| =$ 0.05 rlu, as computed by our
Vaks-type parameterization. The directions [110], [111] and [001] correspond to
the angles 0$^\circ$ or 180$^\circ$, 35.3$^\circ$ or 144.8$^\circ$, and
90$^\circ$, respectively. The experimental data points are shown as open
circle.}
\end{figure}

\begin{figure}[ht]
\begin{center}
\includegraphics{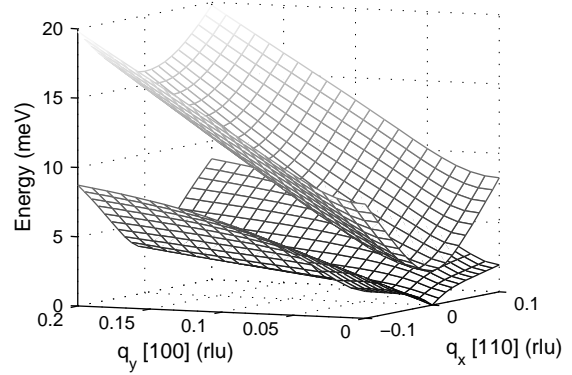}
\end{center}
\caption{\label{nap3d} Highly anisotropic coupled acoustic and optical
phonon surfaces at $T =$ 10 K in KTaO$_3$ showing a deep valley
around the [100] direction; calculated from the modified Vaks parameters.}
\end{figure}

The model provides a global description of the low-frequency spectrum of
KTaO$_3$ as illustrated in Fig. \ref{rot2d} and \ref{nap3d}. A remarkable feature of the phonon
spectrum is the strong anisotropy of the TO and TA modes. We conclude that the \linebreak Vaks model provides a very reasonable
parameterization of the five lowest phonon modes of KTaO$_3$ ($\omega < 8$ meV)
in the central part of the Brillouin zone ($|\vec{q}| < 0.25$ rlu) and in the
low temperature region ($10 < T < 100$ K). The achieved accuracy is $\sim 0.1$ meV in the vicinity of the high symmetry axes, and remains better than $\sim$ 0.2-0.3 meV away from these {\it i.e.} of the order of the experimental uncertainties.

\section{Conclusion}

We presented the results of cold neutron high resolution inelastic scattering
measurements of KTaO$_3$ at low temperatures, for the lowest phonon-dispersion
branches and for wave vectors $|\vec{q}| < 0.25$ rlu. A Monte-Carlo
convolution technique based on a modelisation of the 4-dimensional resolution
ellipsoid was used in order to extract phonon frequencies and dampings from the
experimental spectra. The present results improve significantly on previous
phonon studies in KTaO$_3$ at low-$T$
\cite{Axe70,Shirane67,Harada70,Comes72,Perry89}. In particular, the accuracy on
the phonon frequencies if of the order of 0.1 meV along symmetry directions and
it appears to remain 
better than 0.3 meV in non-symmetry directions. The phenomenological model by
Vaks \cite{Vaks68,Vaks73}, which contains 8 parameters (4 of them being
determined from light-scattering experiments) was used for the parameterization
of the low-energy phonon spectrum. By adjusting 4 temperature-independent
parameters an accurate description of the spectrum was achieved over the
temperature range from 10 to 100 K.

The relatively simple description of the temperature-dependent spectrum of
KTaO$_3$ which is thus obtained provides a convenient basis for quantitative
calculations of all the properties that are controlled by the low frequency
phonons. In particular it becomes possible to calculate thermodynamic and
transport properties at low temperatures, such as phonon-density-fluctuation
processes and second sound, in the low-$T$ quantum paraelectric regime. Such
calculations will be the object of a forthcoming publication.

The authors thank Alain Brochier for his efficient technical support
during the IN14 experiments at ILL.


\newpage



\newpage
\begin{table*}[h] 
\begin{center} 
\begin{tabular}{|c|rrrrrrr|} 
\hline 
$(meV/rlu)^2$ & $S_t$ & $A_t$ & $V_t$ & $S_a$ & $A_a$ & $V_a$ & $A_l$ \\ 
\hline 
\hline 
Ref. \cite{Balagurov70,Vaks73}   & 4860 & 1836 & 2808 & 10800 & 2160 & -5724 & 4320  \\ 
Ref. \cite{Axe70} & 4670 & 1830 & 2800 & & & & \\ 
\hline 
This work    & 4828  & 1553  &  2451 & 8956 &  2265 & -3087 & 4552 \\ 
(uncertainty) & (20)  & (10) & (30) & (400) & (40) & (290) & (30) \\ 
\hline 
\end{tabular} 
\end{center} 
\caption{\label{C3ParFit} Low temperature Vaks parameters of KTaO$_3$,
determined from a global fit of neutron and Brillouin data. See Sec. \ref{sec41effham} for the meaning of the parameters.} 
\end{table*} 
\vspace{1 cm}

\begin{table} 
\begin{center} 
\begin{tabular}{|c|rrrrr|} 
\hline 
$T$ (K) & 80 & 60 & 39 & 24 & 10 \\ 
\hline 
\hline 
$\lambda$ (meV)$^2$&  30.8 &  23.1 &  14.9 &  9.2 &  6.2 \\ 
(uncertainty) &  (0.5) &  (0.8)  & (0.5) &  (0.4)  & (0.3) \\ 
\hline 
\end{tabular} 
\end{center} 
\caption{\label{C3lambda}Square of the soft mode frequency, $\lambda$,
used in our refined Vaks parameterization.} 
\end{table}
\newpage


\end{document}